\RequirePackage{rotating}
\documentclass[fleqn, usenatbib]{mnras}

\usepackage[T1]{fontenc}

\newcommand{\src}{LMC X-4\xspace}
\newcommand{\instrument}{XMM-Newton\xspace}

\DeclareRobustCommand{\VAN}[3]{#2}
\let\VANthebibliography\thebibliography
\def\thebibliography{\DeclareRobustCommand{\VAN}[3]{##3}\VANthebibliography}

\usepackage{graphicx}	
\usepackage{amsmath}	
\usepackage{amssymb}	

\usepackage{flushend}
\usepackage{balance}

\usepackage{float}
\usepackage{caption}
\usepackage{lipsum}

\usepackage{fixltx2e}
\usepackage{xcolor}
\usepackage{tablefootnote}

\usepackage{pdflscape}
\usepackage{longtable}
\usepackage{booktabs}
\usepackage{color}
\usepackage{amssymb}
\usepackage{mathtools}
\usepackage{xspace}
\usepackage{rotating}
\usepackage{appendix}
\usepackage{multirow}
\usepackage{multicol}
\usepackage{array}
\usepackage{subcaption}
\usepackage{comment}
\usepackage{tabularx}
\usepackage{bigstrut}
\usepackage{threeparttable}
\usepackage{afterpage}
\usepackage{capt-of}

\usepackage[normalem]{ulem}

\newcommand{\ecyc}[1]{\ensuremath{E_{\rm{C}}}}

\usepackage{newtxtext, newtxmath}

\title[qpo in lmc]{Discovery of Quasi Periodic Oscillations in the persistent X-ray emission of Accreting Binary X-ray Pulsar \src}

\author[Rikame et al.]{
Ketan Rikame,$^{1,2}$\thanks{E-mail: rikame.bhaskar@res.christuniversity.in}
Biswajit Paul,$^{2}$
Pragati Pradhan,$^{3}$
and KT Paul$^{1}$
\\
$^{1}$CHRIST (Deemed to be University), Department of Physics and Electronics, Bangalore 560029. Karnataka, India\\
$^{2}$Raman Research Institute, Astronomy and Astrophysics, C.V. Raman Avenue, Bangalore 560080. Karnataka, India\\
$^{3}$MIT Kavli Institute for Astrophysics and Space Research, 70 Vassar Street, Cambridge, MA, USA, 02139
}

\date{Accepted XXX. Received YYY; in original form ZZZ}

\pubyear{2021}

\begin{document}
\label{firstpage}
\pagerange{\pageref{firstpage}--\pageref{lastpage}}
\maketitle

\begin{abstract}
We report the discovery of quasi-periodic oscillations (QPOs) in the High Mass X-ray Binary (HMXB) pulsar \src in its non-flaring (persistent) state using observations with XMM-Newton. In addition to the 74 mHz coherent pulsations, the persistent emission light curve shows a QPO feature in the frequency range of 20-30 mHz. Quasi-periodic flares have been previously observed from \src in observations made with Rossi X-ray Timing Explorer (RXTE). However, this is the first time QPOs have been observed in the persistent emission observations of \src. QPOs in X-ray binaries are generally thought to be related to the rotation of the inhomogeneous matter distribution in the inner accretion disk. In High Mass X-ray Binaries (HMXBs) such as \src where the compact object is a neutron star with a high magnetic field, the radius of the inner accretion disk is determined by the mass accretion rate and the magnetic moment of the neutron star. In such systems, the QPO feature, along with the pulse period and X-ray luminosity measurement helps us to constrain the magnetic field strength of the neutron star.
We use considerations of magnetospheric accretion to have an approximate value of the magnetic field strength of the neutron star in \src.
\end{abstract}



\section{Introduction}

\src is a disk fed accretion powered High Mass X-ray binary (HMXB) in the Large Magellanic Cloud first detected with Uhuru \citep{giacconi1972}. This HMXB contains a $1.25_{-0.10}^{+0.11}$ $M_{\odot}$ neutron star and a $14.5_{-1.0}^{+1.1}$ $M_{\odot}$ O8 giant companion \citep{meer2007}. \src is an eclipsing binary pulsar with a pulse period of $\sim$ 13.5 s \citep{white1978,kelly1983} and orbital period of $\sim$ 1.4 day \citep{kelly1983,ilovaisky1984}. Along with the spin period and the orbital motion, a super-orbital variability of 30.4 d is observed \citep{lang1981,molkov2015,paulkitamoto2002}. For an assumed distance of 50 kpc \citep{pietrzy2013}, \src shows unabsorbed continuum X-ray luminosity of $\sim$ 1x10\textsuperscript{38}erg s\textsuperscript{-1} during the high state of the super-orbital period \citep{neilsen2009}, and during the low state, the flux can be lower by a factor of upto 60 \citep{naikpaul2004}. The source exhibits frequent X-ray flares during which the luminosity can reach up-to 10\textsuperscript{39}erg s\textsuperscript{-1} \citep{kelly1983,levine1991,beri2017,brumback2020}.

Accretion disk plays a crucial role in the evolution of accretion-powered HMXBs and it is, therefore, crucial to explore their properties in detail. The instantaneous luminosity allows us to determine the mass accretion rate from the accretion disk. In addition, Quasi Periodic Oscillations (QPOs) observed in the X-ray power spectrum are one of the useful phenomena for studying the interaction of the inner accretion disk with the neutron star magnetosphere. Quasi Periodic Oscillations (QPOs) in X-ray binaries are generally thought to be related to the rotation of in-homogeneous material distribution in the inner accretion disk \citep{ghosh1998}.\\
In the case of black hole X-ray binaries and low magnetic field neutron star X-ray binaries, the accretion disk can extend very close to the compact star. QPOs over a wide range of frequencies from a few Hz to a few hundred Hz are observed in such systems. In accretion powered pulsars such as \src, the strong magnetic field interrupts the disk formation at a large distance. The disk formed at large distances is not hot enough to emit in X-rays. The pulsar beam obstruction due to the accretion disk is also not strong due to the large distance between the neutron star and the accretion disk. Thus strong QPOs are rare in such systems. High magnetic field neutron star systems show only low frequency QPOs \citep{james2010}. We have investigated the timing properties of the accretion powered X-ray pulsar LMC X-4 using observations made with the XMM-Newton and report here the discovery of the QPO feature in its persistent emission state.

\section{Observations, Data Reduction, and analysis}

\src was observed with \instrument six times, including two long duration observations in 2003 (OBSID 0142800101) and 2004 (OBSID 0203500201) and four short duration observations in 2015 (OBSID 0771180101, 0771180201, 0771180301, 0771180401).

\onecolumn

\begin{table}
\begin{center}
\caption{\instrument pointing observation log for \src}
\label{tab:lmcx4_xmm_newton_obs}
\begin{tabular}{c c c c c m{2cm} m{2cm} m{2cm}}
\hline
OBSID	&	Observation Start Date 	&	MJD	&	Exposure (s)	&	$\phi\textsubscript{SO}$	&	\multicolumn{3}{c}{XMM EPIC observation Mode}\\ 
\hline
& & & & & PN & MOS 1 & MOS 2\\
\hline
\hline
142800101	&	2003-09-09	&	52891	&	113653	&	0.92	&	Imaging	&	Imaging	&	Timing\\
203500201	&	2004-06-16	&	53172	&	52896	&	0.16	&	Timing	&	Timing	&	Imaging\\
771180101	&	2015-10-30	&	57325	&	22000	&	0.87	&	Imaging	&	Imaging	&	Timing\\
771180201	&	2015-11-04	&	57330	&	20800	&	0.06	&	Imaging	&	Imaging	&	Timing\\
771180301	&	2015-11-11	&	57337	&	24000	&	0.28	&	Imaging	&	Imaging	&	Timing\\
771180401	&	2015-11-27	&	57353	&	21200	&	0.81	&	Imaging	&	Imaging	&	Timing\\
\hline
$\phi\textsubscript{SO}$: Super-orbital phase\\
\hline
\end{tabular}
\label{obsid}
\end{center}
\end{table}

\begin{figure*}

    \begin{multicols}{2}
    \includegraphics[keepaspectratio=true, height=7.0cm, angle =0]{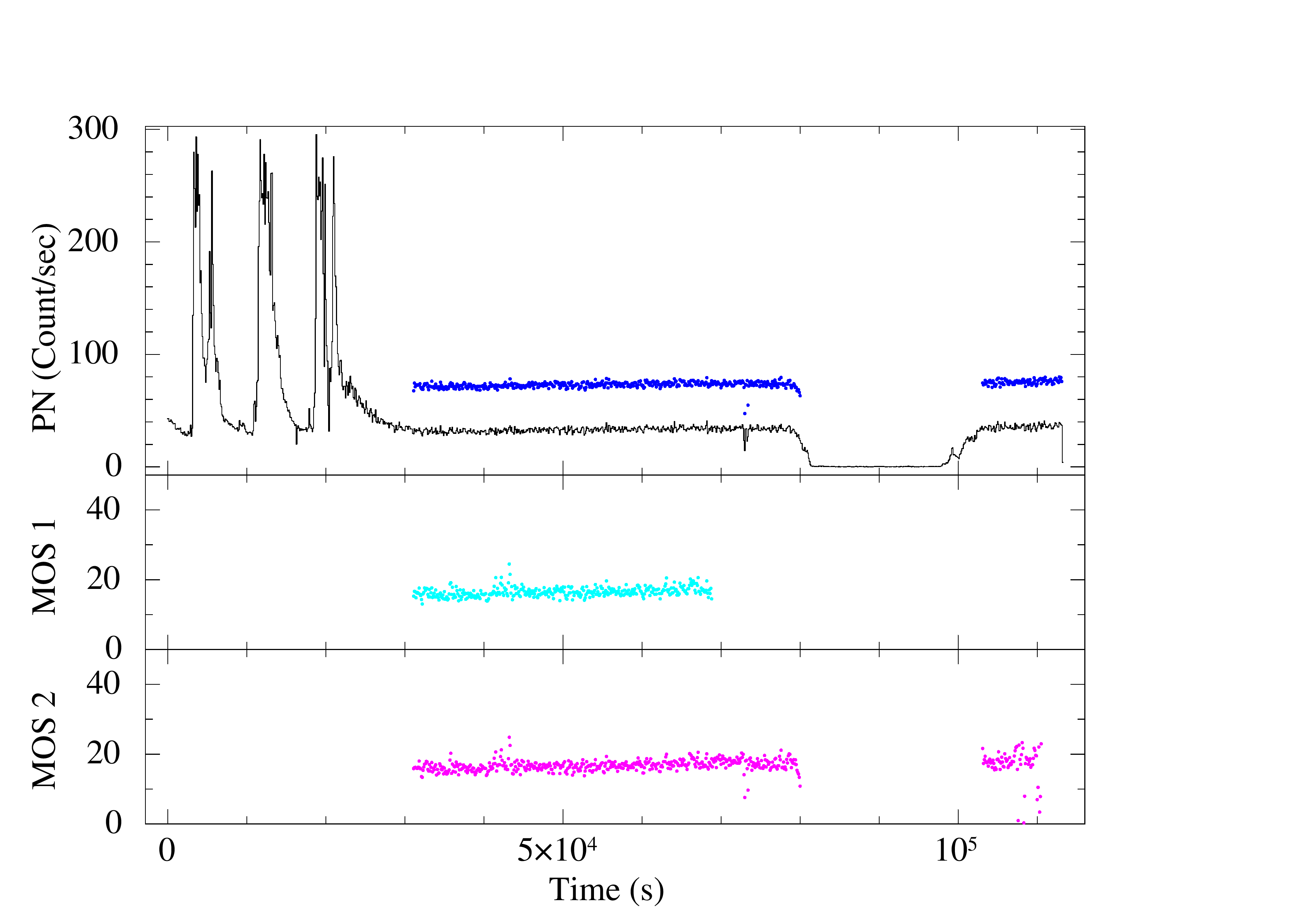}\par
    \includegraphics[keepaspectratio=true, height=7.0cm, angle =0]{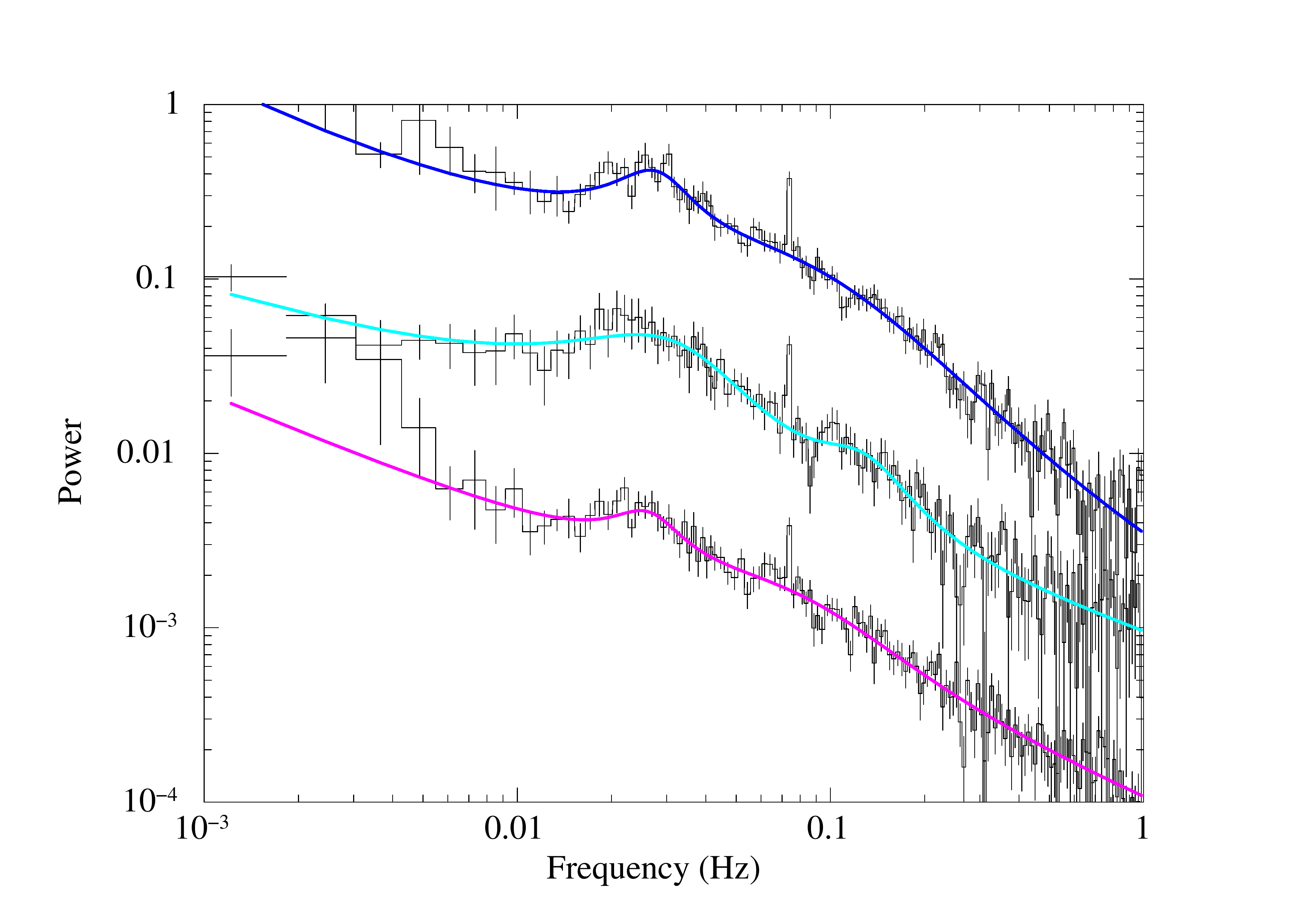}\par
    \end{multicols}
    \caption{\textbf{Left figure:} \instrument EPIC light curve of \src (OBSID: 0142800101). The blue, cyan and magenta colours correspond to the persistent emission PN, MOS1 and MOS2 light curves respectively. Entire duration light curve corresponding to the PN detector is shown in black and an offset of 40 Count/s has been added to the PN detector persistent emission light curve for clarity. \textbf{Right figure:} The PDS corresponding to the persistent emission light curve. MOS 1(2) PDS is multiplied by the factor of 0.1(0.01) for clarity. Thick blue, cyan and magenta lines correspond to the best fitted model for PN, MOS1 and MOS2 light curves respectively. The model comprises of the Power-law and a Lorentzian component to fit the continuum and another Lorentzian component to fit the observed QPO feature. The narrow peak seen on the right of the QPO feature corresponds to the spin frequency of the neutron star.}
    \label{0142800101}

\end{figure*}

\begin{figure*}

    \begin{multicols}{2}
    \includegraphics[keepaspectratio=true, height=7.0cm, angle =0]{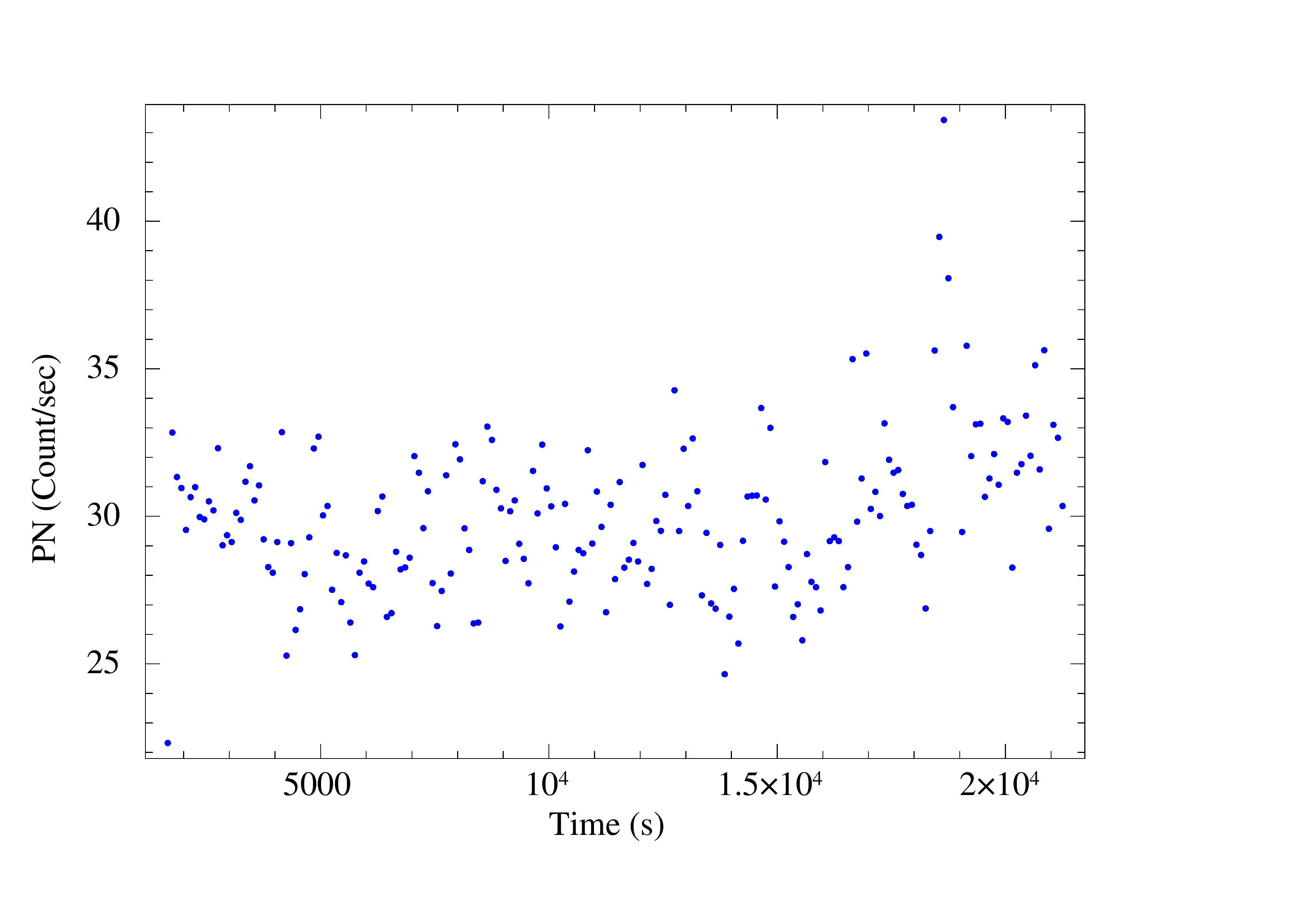}\par
    \includegraphics[keepaspectratio=true, height=7.0cm, angle =0]{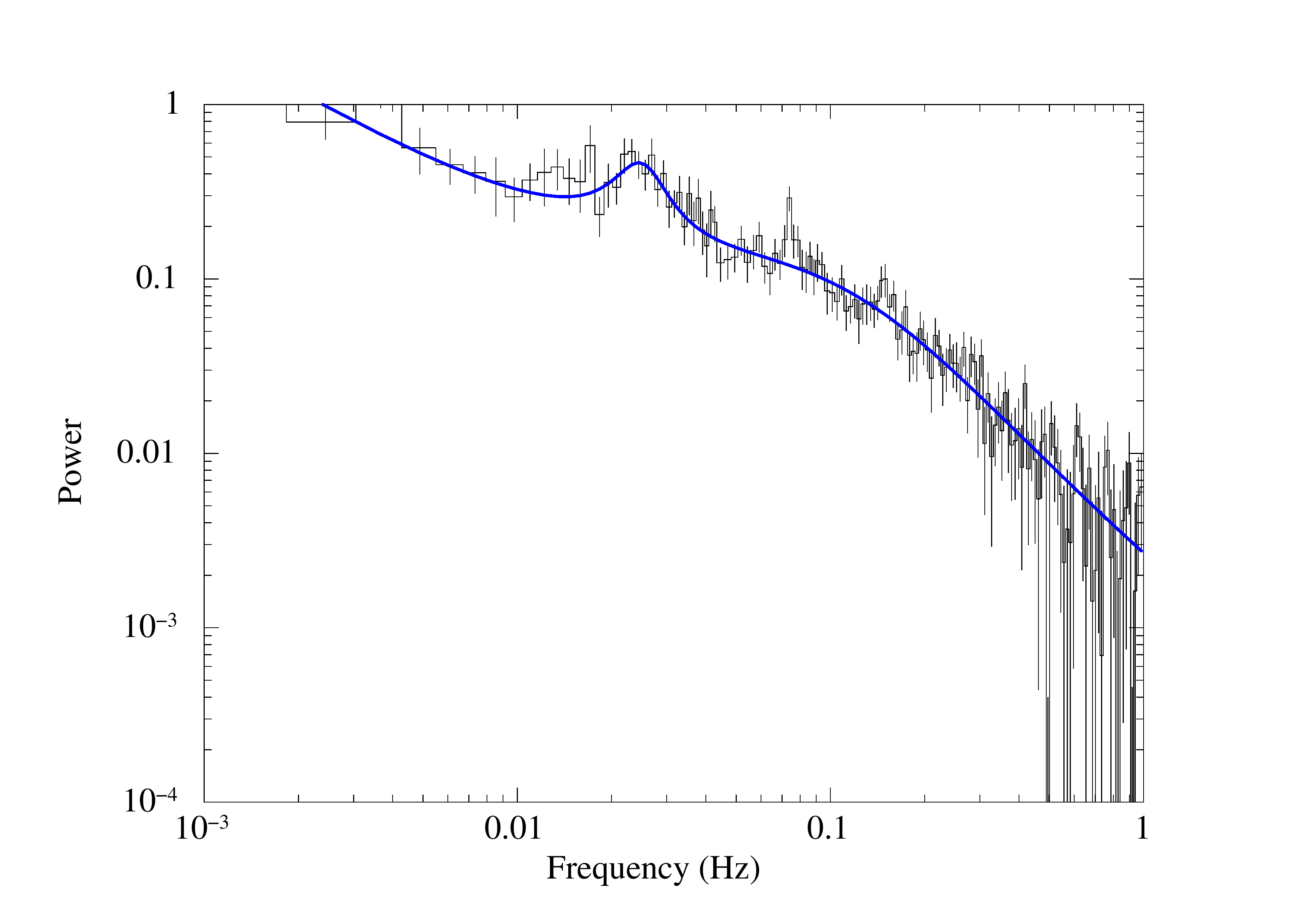}\par
    \end{multicols}
    \caption{\textbf{Left figure:} \instrument EPIC PN entire duration (same as the persistent emission) light curve of \src (OBSID: 0771180401). \textbf{Right figure:} The PDS corresponding to the persistent emission light curve. The blue line represents the best fitted model same as for OBSID 0142800101.}
    \label{0771180401}

\end{figure*}

\twocolumn

These observations have been used earlier to study various aspects of \src \citep{neilsen2009,beri2017,aftab2019,brumback2020}.\\
\instrument has two X-ray instruments aboard, European Photon Imaging Camera (EPIC) and Reflection Grating Spectrometer (RGS). They operate in the energy range of 0.1-15 keV. The EPIC consists of one PN detector \citep{struder_2001} and two MOS detectors \citep{turner_2001}. RGS consists of two spectrometers \citep{herder_2001}. There are three co-aligned telescopes on-board \instrument. The PN detector is at the focal plane of one of the telescopes and receives the complete incoming flux. In other two telescopes, via reflection gratings, part of the flux (44\% of the original incoming flux) reaches the MOS detectors and part of the flux is diverted to Reflection Grating Spectrometers (RGS) \citep{lumb_2012}.\\
The data from PN and both the MOS detectors were used for this work. Table \ref{tab:lmcx4_xmm_newton_obs} shows the log of observations used in this work.\\
Relative super-orbital phase ($\phi\textsubscript{SO}$) is calculated using MJD 53441.50 as epoch ($\phi\textsubscript{SO}$ = 0, peak of the super-orbital intensity modulation) and super-orbital period value of 30.37 days \citep{molkov2015}.\\
Standard Science Analysis System (SAS 18.0.0) was used to process \instrument observation data files (ODF). Procedures given in the online analysis thread were followed (\url{https://www.cosmos.esa.int/web/xmm-newton/sas-threads}). Source data was extracted from a circular region of radius 1 arcmin centered around the source. Background data was extracted from an annular region around the source of inner(outer) radius 1(2) arcmin.
High flaring time intervals and eclipse duration were excluded and only a persistent emission light curve was used for further analysis.
The barycentric correction was applied to the data using the XMMSAS tool \textit{barycen}. Correction for the pulse arrival time delays was implemented using LMC X-4 ephemeris described in \cite{falanga2015}.

\subsection{Timing analysis}
\label{timing analysis}

Persistent emission light curves in the 0.3-10 keV range with a time resolution of 100 ms were generated from all available pointing observations of \src with XMM-SAS. Power density spectrum (PDS) was generated from the source light curves using the FTOOL-powspec \citep{ftools1999}. The light curves were divided into stretches of length 819.2 s and the PDS obtained from each of these segments in one observation were averaged to produce the final PDS. The PDS were normalized such that their integral gives the squared rms fractional variability and the expected white noise level (second-order) was subtracted. The PDS from PN detector data of all six observations and MOS detector data of four of the six observations (OBSID 0142800101, 0203500201, 0771180101, 0771180301), show a peak at around 74 mHz which corresponds to the neutron star spin frequency. In addition to the main peak, a QPO like feature is seen in the frequency range of 20-30 mHz in the PDS of OBSID 0142800101 (PN, MOS1 and MOS2 detector data) and OBSID 0771180401 (PN detector data). Figures \ref{0142800101} and \ref{0771180401} show light curves from PN detector data (left panel) and corresponding PDS plots (right panel) for OBSID 0142800101 and 0771180401 respectively. For OBSID 014288101, the persistent emission light curve (shown in green) has been offset by 40 Counts/s from the entire duration light curve for clarity. No offset is required for the light curve of the OBSID 0771180401 since the entire observation is acquired during the persistent emission state and no data is excluded.\\
To obtain the peak frequency of the QPO feature, the PDS was fit with two continuum components, a power-law, and a Lorentzian, with one more Lorentzian component to fit the QPO feature when present. The peak corresponding to the neutron star spin frequency is avoided during fitting. The thick blue, cyan and magenta lines in the PDS plot shows the best fit model for PN, MOS1 and MOS2 light curves respectively.\\
A prominent QPO feature is observed in the persistent emission data from the PN detector of the longest duration observation acquired in the year 2003 (OBSID 0142800101). The best PDS fit obtained a reduced $\chi\textsuperscript{2}$ of 1.16 for 138 degrees of freedom. The QPO frequency for this observation was measured to be around 27 mHz with rms fraction of 7.6\% and a quality factor (QPO frequency/FWHM) of 1.50.\\
The energy dependence of the QPO feature was also investigated for this data. The power spectra in the low energy band (0.3-1 keV) and high energy band (1-10 keV) were observed. The energy bands were chosen such that there are a similar number of photons in both bands. Also as will be evident from the spectral analysis, the thermal components are dominant up to $\sim$ 1 keV and non-thermal components dominate beyond 1 keV in the X-ray spectrum. The pulsation peak is observed to be more pronounced in the low energy band. The low energy power spectral fit obtained a reduced $\chi\textsuperscript{2}$ of 1.14 for 138 degrees of freedom and the QPO frequency of $25.3_{-1.4}^{+1.4}$ with a quality factor of 1.31. The same for high energy band obtained a reduced $\chi\textsuperscript{2}$ of 1.86 for 138 degrees of freedom and the QPO frequency of $27.9_{-1.8}^{+1.9}$ mHz with a quality factor of 2.27. The rms variability in the low energy band is 10.1 \% and in the high energy band is 5.3\%. The characteristics of the QPO observed in the PDS of the rest of the observations are listed in the Table \ref{tab:pspec_para}\\

\begin{table}
\begin{center}
\caption{Characteristics of the QPO observed in the PDS of persistent emission data of \src}
\label{tab:pspec_para}
\begin{tabular}{ c m{1.8cm} m{1.1cm} m{1cm} m{1cm} }
\hline
OBSID   & QPO frequency &   RMS variablity    &   Quality factor\textsuperscript{a}  & Fit Statistics\textsuperscript{b}\\
& (mHz) &   (\%)    &   &\\
\hline
0142800101 &&&&\\
\hline
\\[-1em]
PN  &  $26.8_{-1.4}^{+1.5}$ &   7.6 &   1.50    &   1.16/138\\
\\[-1em]
MOS1  &  $25.5_{-3.7}^{+3.2}$ &   14.0 &   0.60    &   1.22/138\\
\\[-1em]
MOS2  &  $25.5_{-2.6}^{+2.1}$ &   6.7 &   1.66    &   1.07/138\\
\\[-1em]
\hline
0771180401 &&&&\\
\hline
\\[-1em]
PN  &  $24.1_{-1.3}^{+1.3}$ &   6.2 &   2.45    &   0.99/138\\
\\[-1em]
\hline
\multicolumn{5}{c}{a: Quality factor = QPO frequency / FWHM}\\
\multicolumn{5}{c}{b: Reduced $\chi$\textsuperscript{2} / Degrees of freedom (Fit statistics format)}\\
\hline
\end{tabular}
\label{pspec_para}
\end{center}
\end{table}

\subsection{Spectral analysis}
\label{spectral analysis}

\begin{figure}
    \includegraphics[keepaspectratio=true, height=7.0cm, angle =0]{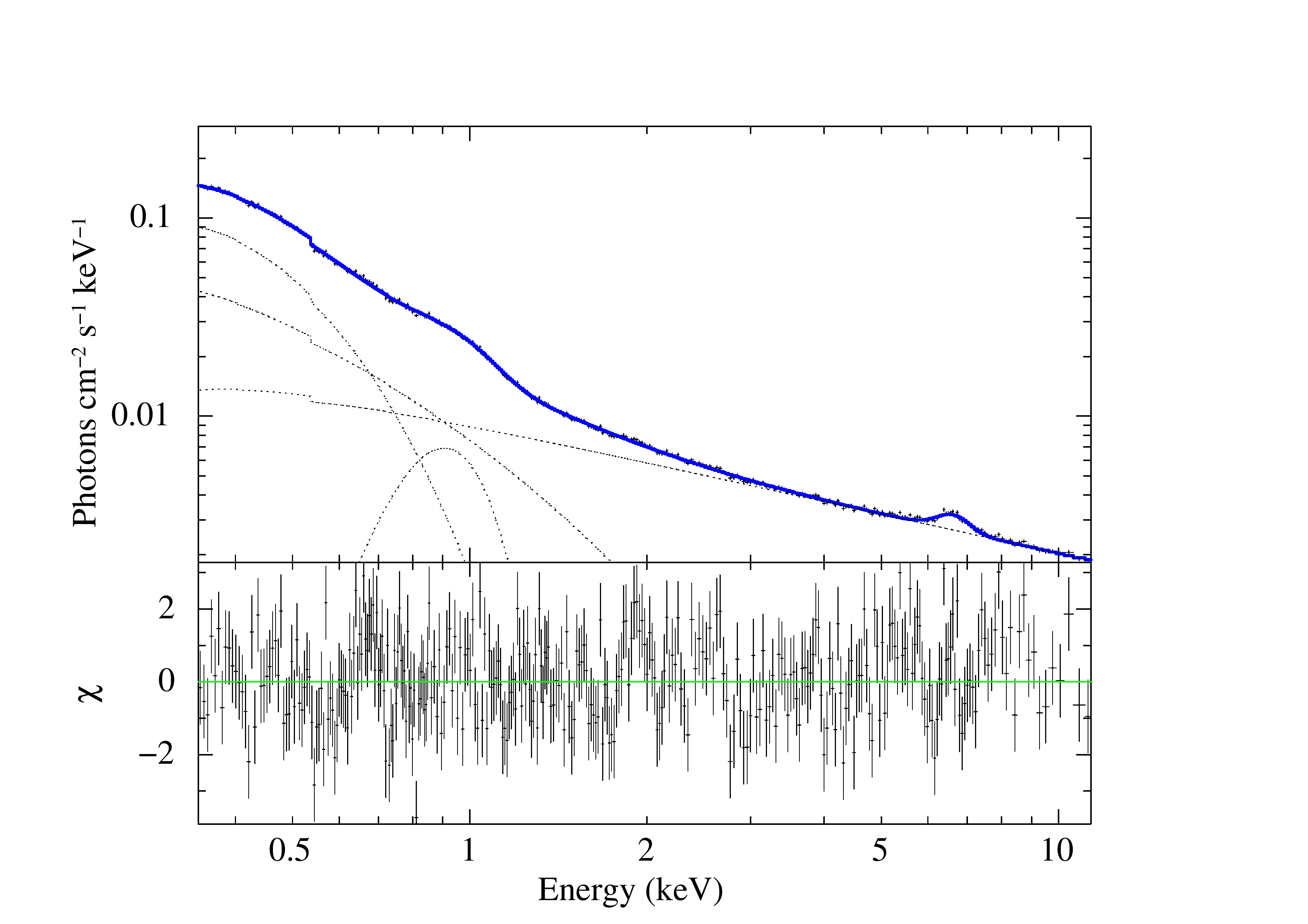}
    \caption{The X-ray spectrum of persistent emission data from LMC X–4 (OBSID 0142800101). The solid blue line shows the best fit spectral model fitted with a power-law component to fit for the continuum, bremsstrahlung, and a blackbody component to fit for the soft excess and two Gaussian components to fit for the emission features observed near 1 keV and 6.4 keV. The spectrum is attenuated with a line of sight absorption. The dotted lines show the individual model components. The bottom panel shows the contributions of the residuals to the chisq.}
    \label{0142800101_spec}
\end{figure}

\begin{figure}
    \includegraphics[keepaspectratio=true, height=7.0cm, angle =0]{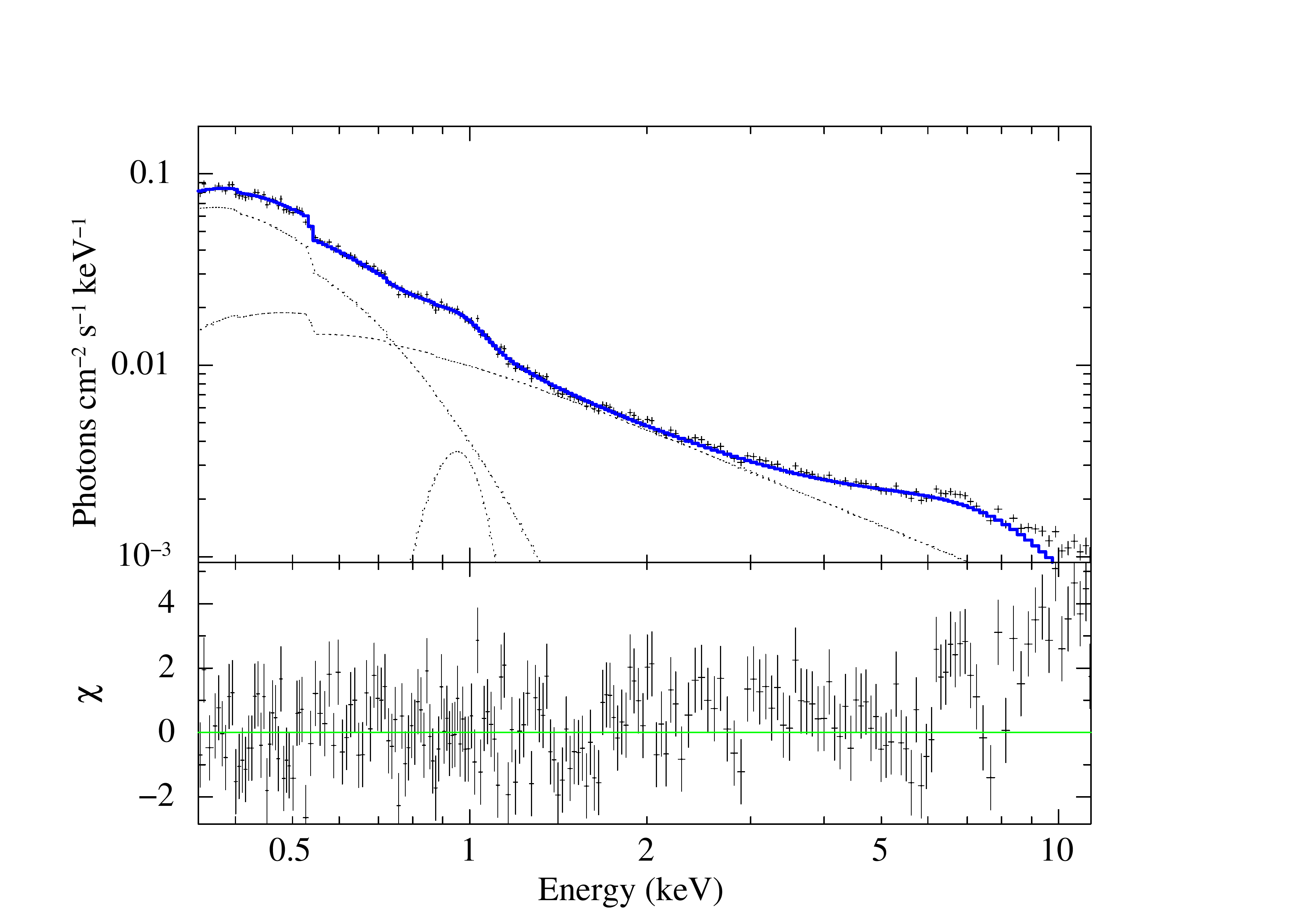}
    \caption{Same as Figure \ref{0142800101_spec} for OBSID 0771180401.}
    \label{0771180401_spec}
\end{figure}

Spectral analysis was performed on the persistent emission data of the observations in which the QPO feature was observed (PN detector data of OBSID 0142800101, 0771180401). XMM-SAS tool \textit{epatplot} showed presence of pile-up in both the observations. In the case of OBSID 0142800101, central part of radius 7.5 arcseconds was removed from the source region to mitigate the effects of pile-up. In the case of OBSID 0771180401, central part of radius 13 arcseconds was removed from the source region for the same reason. The measures taken to remove the pile-up are in agreement with the ones reported in \cite{beri2017} and \cite{brumback2020}.\\
Analysis was performed using \textit{Xspec Version}: 12.11.0 \citep{xspec1996}. Similar to \cite{beri2017}, the mean spectrum is extracted and fitted with a power-law component to fit for the continuum, bremsstrahlung, and a blackbody component to fit for the soft excess and two Gaussian components to fit for the emission features observed near 1 keV and 6.4 keV. The blackbody component is not used while fitting the spectrum corresponding to the persistent emission data of OBSID 0771180401, since the spectral parameters of blackbody model are insensitive to the spectral fitting. The spectrum is attenuated with a line of sight absorption. Table \ref{tab:spec_para} lists the spectral parameters and their uncertainties.\\
X-ray spectra are shown in figures \ref{0142800101_spec} and \ref{0771180401_spec}. The solid blue line shows the best fit spectral model and the dotted lines show the individual model components. The contributions of the residuals to the chisq are shown in the bottom panel.\\
X-ray flux uncertainties for an obtained best-fit spectral model are negligible (Ref. Table \ref{tab:spec_para}) and thus not considered for further analysis. The unabsorbed X-ray flux is obtained by correcting for the attenuation due to line of sight absorption. The spectral fit corresponding to the OBSID 0142800101 obtained a reduced $\chi\textsuperscript{2}$ value of 1.08 for 2204 degrees of freedom and the unabsorbed X-ray flux in the 0.3-10 keV band is 2.97 x 10\textsuperscript{-10} erg s\textsuperscript{-1} cm\textsuperscript{-2}. The reduced $\chi\textsuperscript{2}$ value is 1.12 for 2204 degrees of freedom in the case of OBSID 0771180401 and the unabsorbed X-ray flux in the 0.3-10 keV band is 2.06 x 10\textsuperscript{-10} erg s\textsuperscript{-1} cm\textsuperscript{-2}.\\
Along with the QPO frequency, the luminosity measurement is required to obtain an estimate of the magnetic field strength of the Neutron Star, hence the spectral analysis is performed.

\begin{table}
\begin{center}
\caption{Best-fitting parameters for the pile-up corrected spectrum of \src}
\label{tab:spec_para}
\begin{tabular}{ m{3cm} c c }
\hline
Parameters	&	\multicolumn{2}{c}{OBSID}\\ 
\hline
& 0142800101 & 0771180401\\
\hline
\hline
N\textsubscript{H} (10\textsuperscript{22} atoms cm\textsuperscript{-2})&(1.65$\pm$0.35) x 10\textsuperscript{-2}&(5.70$\pm$0.65) x 10\textsuperscript{-2}\\
Photon Index ($\Gamma$)&0.66$\pm$0.01&1.28$\pm$0.09\\
Norm\textsubscript{\textit{power-law}}&(0.92$\pm$0.02) x 10\textsuperscript{-2}&(1.14$\pm$0.05) x 10\textsuperscript{-2}\\
$kT\textsubscript{\textit{bremss}}$ (keV)&(10.47$\pm$0.90) x 10\textsuperscript{-1}&(2.76$\pm$0.17) x 10\textsuperscript{-1}\\
Norm\textsubscript{\textit{bremss}}&(1.52$\pm$0.12) x 10\textsuperscript{-2}&(8.65$\pm$1.19) x 10\textsuperscript{-2}\\
$kT\textsubscript{\textit{bbodyrad}}$ (keV)&(1.02$\pm$0.02) x 10\textsuperscript{-1}&-\\
Norm\textsubscript{\textit{bbodyrad}}&(0.04$\pm$0.00) x 10\textsuperscript{-2}&-\\
Energy\textsubscript{\textit{Gaussian 01}} (keV)&(9.02$\pm$0.08) x 10\textsuperscript{-1}&(9.47$\pm$0.10) x 10\textsuperscript{-1}\\
Width\textsubscript{\textit{Gaussian 01}} (keV)&(1.58$\pm$0.07) x 10\textsuperscript{-1}&(0.96$\pm$0.12) x 10\textsuperscript{-1}\\
Norm\textsubscript{\textit{Gaussian 01}}&(0.29$\pm$0.02) x 10\textsuperscript{-2}&(0.10$\pm$0.01) x 10\textsuperscript{-2}\\
Energy\textsubscript{\textit{Gaussian 02}} (keV)&6.62$\pm$0.02&6.28$\pm$0.14\\
Width\textsubscript{\textit{Gaussian 02}} (keV)&0.42$\pm$0.02&2.44$\pm$0.31\\
Norm\textsubscript{\textit{Gaussian 02}}&(0.05$\pm$0.00) x 10\textsuperscript{-2}&(0.57$\pm$0.15) x 10\textsuperscript{-2}\\
\hline
0.3-10 keV model flux (10\textsuperscript{-10} ergs cm\textsuperscript{-2} s\textsuperscript{-1})	&	$2.89_{-0.01}^{+0.00}$	&	$1.86_{-0.01}^{+0.01}$\\
\hline
Reduced $\chi$\textsuperscript{2} / Degrees of freedom	&	1.08/2204	&	1.12/2204\\
\hline
\hline
\multicolumn{3}{c}{Model: phabs * (bremss + powerlaw + gaussian + gaussian)}\\
\hline
\end{tabular}
\label{spec_para}
\end{center}
\end{table}

\section{Discussion}
\label{discussion}

A QPO feature of transient nature has been observed in the \instrument persistent emission data of accreting HMXB pulsar \src. Two of the six available \instrument pointing observations exhibit QPO feature in the persistent emission data. \cite{moon2001} have previously observed milli-hertz (mHz) quasi-periodic flares and related changes in the pulsations during large X-ray flares from \src using the data from Rossi X-ray Timing Explorer (RXTE). However, this is the first time QPOs have been detected in the persistent emission observations of \src.

QPOs in the range 1mHz-20Hz have been observed in several HMXB pulsars, with persistent and transient X–ray emission \citep{finger1998}, both with Be and early type supergiant companions (see \cite{james2010} for a list of sources). For most of the accretion powered pulsars, the QPOs are a transient phenomenon. QPOs are also present in accreting millisecond pulsars (AMXPs) at a few hundred hertz \citep{vanderklis_lmxb_qpo_1998}.

Several models have been proposed to explain the origin of the QPO feature in the accretion powered X-ray pulsars among which the Keplerian frequency model (KFM) and Magnetospheric beat frequency model (MBFM) are used most commonly.
According to the KFM, QPOs arise from the modulation of the X-rays by inhomogeneities in the inner disk at the Keplerian frequency \citep{kfm_vanderklis_1987}. Here, the QPO frequency is the Keplerian orbital frequency of inhomogeneities in the inner accretion disk. The observed QPO frequency is around 25 mHz which is about 1/3\textsuperscript{rd} of the spin frequency of the neutron star. If the neutron star is spinning at a faster rate than the inner accretion disk (as is the case if KFM is considered), the infalling material will gain angular momentum and will be blown away due to centrifugal inhibition. Thus, for successful matter accretion, the neutron star spin frequency has to be slower than the orbital frequency of the inner accretion disk. This condition can be satisfied by assuming MBFM.
According to the MBFM, QPOs occur at the beat frequency between the orbital frequency of matter in the accretion disk at the Alfv´en radius (the distance where magnetic pressure is equal to the ram pressure of the infalling matter) and the stellar spin frequency \citep{bfm_shaham_1987}.
In the case of prominent QPO observed in OBSID 0142800101 (longest duration observation acquired in the year 2003), the Keplarian frequency of the inner accretion disk as per MBFM is $\nu\textsubscript{k} = \nu\textsubscript{QPO} + \nu\textsubscript{s}$ = (74 + 26) mHz = 100 mHz whereas the same in the case of QPO observed in OBSID 0771180401 (acquired in the year 2015) is (74 + 24) mHz = 98 mHz. The radius of the inner accretion disk can be estimated as,

     \begin{align*}
         R\textsubscript{MBFM} = \left(\frac{GM}{4\pi^2(\nu\textsubscript{QPO} + \nu\textsubscript{s})^2}\right)^\textsuperscript{1/3}.
     \end{align*}

The R\textsubscript{MBFM} obtained for a neutron star mass of 1.4$M_{\odot}$ is 7.8 x 10\textsuperscript{3} km for OBSID 0142800101 and 7.9 x 10\textsuperscript{3} km for OBSID 0771180401.\\
The X-ray flux was measured from the best fit spectral model of the persistent emission data of both the observations in which QPO feature was observed. The unabsorbed X-ray flux in the 0.3-10 keV band is measured to be 2.97 x 10\textsuperscript{-10} erg s\textsuperscript{-1} cm\textsuperscript{-2} for the OBSID 0142800101 and 2.06 x 10\textsuperscript{-10} erg s\textsuperscript{-1} cm\textsuperscript{-2} for the OBSID 0771180401. The luminosity was calculated from the X-ray flux using the assumed source distance of 50 kpc \citep{pietrzy2013}. The X-ray luminosity of 8.89 x 10\textsuperscript{37} erg s\textsuperscript{-1} and 6.16 x 10\textsuperscript{37} erg s\textsuperscript{-1} was obtained for the OBSID 0142800101 and OBSID 0771180401 respectively.
Along with spin and orbital period variability, \src shows super-orbital variability of around 30.4 days \citep{lang1981,molkov2015,paulkitamoto2002}. While in some systems it is argued that the superorbital variability may be caused by some intrinsic changes in the X-ray flux \citep{pradhan2020}, in the case of LMC X-4, the superorbital variability is known to be caused by changes in absorption due to motion of titled/warped accretion disk. The X-ray luminosity in the high state of the super-orbital period is closest to the intrinsic luminosity of the source. The time duration of the OBSID 0142800101 falls in the high state ($\phi\textsubscript{SO}$ = 0.92) of the super-orbital period. Naturally, the X-ray luminosity of this observation i.e. 8.89 x 10\textsuperscript{37} erg s\textsuperscript{-1} is also higher among the two observations showing the QPO feature. This value of X-ray luminosity is used as the intrinsic luminosity for further analysis.

Using the considerations of magnetospheric accretion, the radius of the magnetosphere of a neutron star can be expressed in terms of the luminosity and magnetic moment as \citep{frank2002},

     \begin{align*}
         R\textsubscript{M} = 2.9 \cdot 10^8 \cdot m\textsubscript{1}\textsuperscript{1/7} \cdot R\textsubscript{6}\textsuperscript{-2/7} \cdot L\textsubscript{37}\textsuperscript{-2/7} \cdot \mu\textsubscript{30}\textsuperscript{4/7} cm
     \end{align*}

where m\textsubscript{1} is the neutron star mass in the units of 1.4$M_{\odot}$, R\textsubscript{6} is the radius of the neutron star in units of 10\textsuperscript{6} cm, L\textsubscript{37} is the X-ray luminosity in units of 10\textsuperscript{37} erg s\textsuperscript{-1} and $\mu\textsubscript{30}$ is the magnetic moment in units of 10\textsuperscript{30} Gauss cm\textsuperscript{3}.
In the case of high magnetic field neutron stars such as the one in \src, the magnetic field starts to dominate near the stellar surface. The accreting material is channeled along the magnetic field lines onto the magnetic poles disrupting the accretion disk inside the magnetosphere. As the accretion disk can't sustain inside the magnetosphere, R\textsubscript{M} and R\textsubscript{MBFM} can be equated and the magnetic moment can be estimated. Considering $\nu\textsubscript{k}$ = 100 mHz and using larger one of the two fluxes measured with XMM-Newton, the magnetic moment is estimated to be 16.76 x 10\textsuperscript{30} Gauss cm\textsuperscript{3}.
Other measurements of X-ray luminosity of LMC X-4 during the peak of the superorbital period are 2.3 x 10\textsuperscript{38} erg s\textsuperscript{-1} measured with RXTE in 2-25 keV band \citep{levine2000}, and 3.0 x 10\textsuperscript{38} erg s\textsuperscript{-1} measured with Beppo-SAX in the 0.1-100 keV band \citep{labarbera2001}. Assuming a mass of 1.4$M_{\odot}$ and a radius of 10 km for the neutron star, these luminosity values give a magnetic moment of 26.97 x 10\textsuperscript{30} Gauss cm\textsuperscript{3} and 30.80 x 10\textsuperscript{30} Gauss cm\textsuperscript{3} respectively for the neutron star. For a dipole-like magnetic field, the field strength B varies roughly as $\mu/R\textsuperscript{3}$. Thus the magnetic field strength on the neutron star surface (i.e. at $R$ = 10\textsuperscript{6} cm) is estimated to be approximately around 30.80 x 10\textsuperscript{12} Gauss for $\nu\textsubscript{k}$ = 100 mHz and using highest of the persistent luminosities measured till date \citep{labarbera2001}. We also note that in calculation of the magnetospheric radius, there is a correction factor of the order of unity due to disk accretion and in determination of the surface magnetic field, there is uncertainty about the radius of the neutron star. The cyclotron resonant scattering feature (CRSF) in the X-ray spectrum is the only direct way to measure the neutron star’s magnetic field \citep{truemper1978}. In spite of all the uncertainties, the CRSF feature in LMC X-4 is expected to be observed at more than 100 keV. Evidence of a broad absorption feature at 100 keV was found in the same Beppo-SAX observation \citep{labarbera2001}, though the parameters of the features could not be constrained well as it was at the high energy limit of the telescope.\\
Several parameters and considerations mentioned in the discussion above may change under different circumstances and alter the magnetic field estimate. The highest persistent emission luminosity estimate is considered to be the intrinsic luminosity of the source however the actual intrinsic luminosity can be different due to the beaming of the pulse averaged X-rays in different directions. R\textsubscript{M} and R\textsubscript{MBFM} are equated considering that the disruption of the accretion disk to channel the accreting matter along the magnetic field lines onto the magnetic poles happens at the magnetospheric boundary. The azimuthal component of the magnetic torque responsible for disruption of accretion disk varies with the orientation of the magnetic field. Depending upon the magnetic field orientation under consideration, the disruption of the accretion disk can happen inside or outside the magnetosphere \citep{frank2002}. QPOs alone do not provide sufficient information regarding these. However, the expression for the magnetic field is relatively insensitive to these parameters, and variation in these parameters does not change the final estimate of the magnetic field strength by much. The magnetic field expression has a very steep dependence on the radius of the neutron star though. The neutron star radius is assumed to be 10 km and varying it by 10\% varies the magnetic field estimate by 25\%.\\
The continuous accretion of matter onto the neutron star induces accretion torque resulting in the spinning up of the neutron star. Such spin-up has been observed in systems like GRO J1744-28, A0535+262, etc. \citep{bildsten1997}. Occasional spin-down has also been observed in systems like Cen X-3 which can be attributed to the fluctuations in the accretion torque, change in the internal structure of the neutron star, etc. The net result however is a slow spin-up \citep{bildsten1997}. Also, if the neutron star is spun up to rotate at a faster rate than the accretion disk, the matter accretion onto the neutron will be prohibited due to centrifugal inhibition. Thus eventually neutron star spin period (P\textsubscript{spin}) would attain an equilibrium. Using the considerations of magnetospheric accretion, the equilibrium spin period (P\textsubscript{eq}) can be estimated.

     \begin{align*}
         P\textsubscript{eq} \sim 3 \cdot m\textsubscript{1}\textsuperscript{-2/7} \cdot R\textsubscript{6}\textsuperscript{-3/7} \cdot L\textsubscript{37}\textsuperscript{-3/7} \cdot \mu\textsubscript{30}\textsuperscript{6/7} s
     \end{align*}

where the notations are the same as mentioned previously \citep{frank2002}. The system would remain in equilibrium until external conditions such as accretion rate change. Using the approximate estimation of the magnetic moment of the neutron star in \src, the equilibrium spin period of the neutron star in \src is estimated to be around 13.18 s, close to its current spin period of about 13.5 s. Thus \src appears to be close to its equilibrium spin period.

\section{Conclusion}
\label{conclusion}
A QPO feature of transient nature has been discovered at around 25 mHz in the persistent emission data of accreting HMXB pulsar \src. Using the X-ray luminosity and the measured QPO frequency and using the considerations of magnetospheric accretion, an approximate value of the magnetic field strength of the neutron star in \src has been estimated.

\section*{Acknowledgements}
This research has made use of archival data and software provided by NASA's High Energy (HEASARC), which is a service of the Astrophysics Science Division at NASA/GSFC.
We thank an anonymous referee for the valuable comments that helped us improve the paper.

\section*{Data Availability}
The observational data underlying this work is publicly available through the High Energy Astrophysics Science Archive Research Center (HEASARC). Any additional information will be shared on reasonable request to the corresponding author.



\bibliographystyle{mnras}
\bibliography{reference_lmc} 



\bsp	
\label{lastpage}
\end{document}